\documentclass[aps,prb,preprint]{revtex4-1}

\usepackage{graphicx}
\usepackage{amsmath}
\usepackage{textcomp} 

\usepackage{color}

\begin{document}

\title{Creation of uni-directional spin-wave emitters by utilizing interfacial Dzyaloshinskii-Moriya interaction}

\author{T. Br\"{a}cher}
\email{thomas.braecher@cea.fr}
\affiliation{SPINTEC, UMR-8191, CEA-INAC/CNRS/UJF-Grenoble/Grenoble-INP, 17 Rue des Martyrs, 38054 Grenoble, France}
\author{P. Pirro}
\affiliation{Fachbereich Physik and Landesforschungszentrum OPTIMAS, Technische Universit\"at Kaiserslautern, 67663 Kaiserslautern, Germany}
\author{O. Boulle}
\affiliation{SPINTEC, UMR-8191, CEA-INAC/CNRS/UJF-Grenoble/Grenoble-INP, 17 Rue des Martyrs, 38054 Grenoble, France}
\author{G. Gaudin}
\affiliation{SPINTEC, UMR-8191, CEA-INAC/CNRS/UJF-Grenoble/Grenoble-INP, 17 Rue des Martyrs, 38054 Grenoble, France}

\date{\today}

\begin{abstract}

We present an analytic and numerical study of the creation of an uni-directional spin-wave emission in ultra-thin ferromagnetic films sandwiched in an asymmetric layer stack. For this we extend the analytical description of spin waves in spin-wave waveguides by the incorporation of the influence of the interfacial Dzyaloshinskii-Moriya interaction on the spin-wave propagation. By exploring the model system Ni$_{81}$Fe$_{19}$/Pt we show that it is possible to achieve an uni-directional spin-wave emission by combining wave-vector selective excitation sources with the frequency splitting which arises from interfacial Dzyaloshinskii-Moriya interaction. Hereby we focus on device feature sizes and spin-wave wavelengths compatible with state-of-the art excitation and detection schemes. We demonstrate that the optimum operation conditions for the non-reciprocal emission can be predicted using the presented analytical formalism.

\end{abstract}

\pacs{}

\maketitle

\section{Introduction}
\label{Intro}

The research field of magnonics\cite{Magnon-Spintronics,Neusser,Lenk-2011-1,Magnonics-Krug,Davies-2015-1,TAP-2013} addresses the transport of information in the form of spin waves and the fundamental study of the spin-wave properties in magnetic layer systems. Recently, the spin-wave dynamics in thin magnetic layers sandwiched in an asymmetric layer stack have attracted an increased research interest. This is motivated by the large variety of spin-orbit coupling phenomena which appear in such layer systems with broken structural inversion symmetry\cite{Sampaio-2013-1,Manchon-2014-1,Miron-2010-1, Sklenar-2016-1}. Besides the consequent current-induced effects on the spin-wave dynamics, like the current-induced modulation of the spin-wave damping\cite{Gladii-2016-2, MockelPyPt,Demi-2014-STNO}, this structural inversion asymmetry also influences the spin-wave properties in the absence of a (charge) current. One of these effects is the frequency splitting of the spin-wave dispersion for spin waves traveling in the magnetostatic surface wave configuration, i.e., waves traveling perpendicular to the direction of the static magnetization in an in-plane magnetized ferromagnetic film\cite{DE, Stancil}. In addition to asymmetric interface anisotropies\cite{Burkard-PRL-1990,Gladii-2016} in thin films with thickness of a few tens of nanometers, interfacial Dzyaloshinskii-Moriya interaction (iDMI)\cite{Fert-1990, Di-2015, Moon-2013, Kostylev-2014, Nembach-2015,Stashkevich-2015,Lee-2016} has been recently identified as the origin of such frequency splittings in ultra-thin films (thicknesses of a few nanometers and below). In turn, spin waves have been established as a sensitive probe of the interface characteristics and, next to the study of the domain wall motion\cite{Boulle-2013,Martinez-2013,Brataas-2013,Emori-2013,Ryu-2013,Hrabec-2014-1,Je2013}, they have become an indispensable probe for iDMI. 

Recent studies also concerned the effect of the iDMI on the spin-wave spectrum and its potential application. In particular, the realization of a diode-like, uni-directional spin-wave emission was addressed in literature\cite{Lan-2015, Sanchez-2014}, since it promises an efficient control of the spin-wave emission in spin-wave based applications\cite{Magnon-Spintronics}. These previous studies were focused on the utilization of exchange-dominated spin waves, which, due to their large wave-vectors, feature a sizable iDMI-induced splitting of the dispersion. However, these waves generally suffer from a short spin-wave lifetime and they are difficult to access in standard experiments such as Brillouin light scattering spectroscopy or all-electrical propagating spin-wave spectroscopy\cite{BLS1,BLS2,BLS3,Bailleul-2003,Florin-2016,Koji-2012,Koji-2014}. In contrast, the spectrum of dipole-exchange spin waves lies in a wave-vector range compatible with standard detection and excitation schemes\cite{Mockelantenne,Pirro-2011-1}. Here, we provide an analytical and numerical study of the realization of a uni-directional emission of dipole-exchange spin waves in spin-wave waveguides featuring only a weak iDMI. We demonstrate that the wave-vector selectivity provided by a coplanar waveguide (CPW) can suffice to achieve this uni-directionality. By applying the theory for spin waves in thin films to microscopic magnonic waveguides featuring iDMI, we give guidelines for the practical realization of these uni-directional emitters. As model system we choose a layer system incorporating Ni$_{81}$Fe$_{19}$ (Permalloy, Py) and Pt, a system which is commonly used in magnon spintronic applications\cite{Magnon-Spintronics,Neusser,Lenk-2011-1,Magnonics-Krug,Davies-2015-1,TAP-2013, MockelPyPt,Gladii-2016-2,Demi-2014-STNO}. Recently, a weak, yet sizable iDMI was found in this material system\cite{Nembach-2015,Stashkevich-2015}, which is about one order of magnitude lower than the iDMI in Pt/Co/Al$_2$O$_3$\cite{Belmeguenai-PRB-2015} or Pt/Co/MgO\cite{Boulle-2015-RTS}. These systems are known to exhibit a large iDMI, but they feature a significantly larger intrinsic damping.

The structure of the article is as follows: In Section \ref{Anal} we present an overview of the analytical description of spin waves in waveguides made from thin ferromagnetic films and the incorporation of iDMI into this description. Consequently, we present the conducted numerical study of the uni-directional emission in the model system Py/Pt in Section \ref{Num}. Thereafter, we compare the analytical model to the numerical simulations in Section \ref{Res} and we demonstrate the realization of the uni-directional spin-wave emission in Section \ref{Uni}. Ultimately, the paper closes with a discussion of the obtained results in Section \ref{Disc} and a conclusive statement in Section \ref{Conc}.

\section{Analytical model}
\label{Anal}

\begin{figure*}[t!]
	  \begin{center}
    \scalebox{1}{\includegraphics[width=0.9\linewidth, clip]{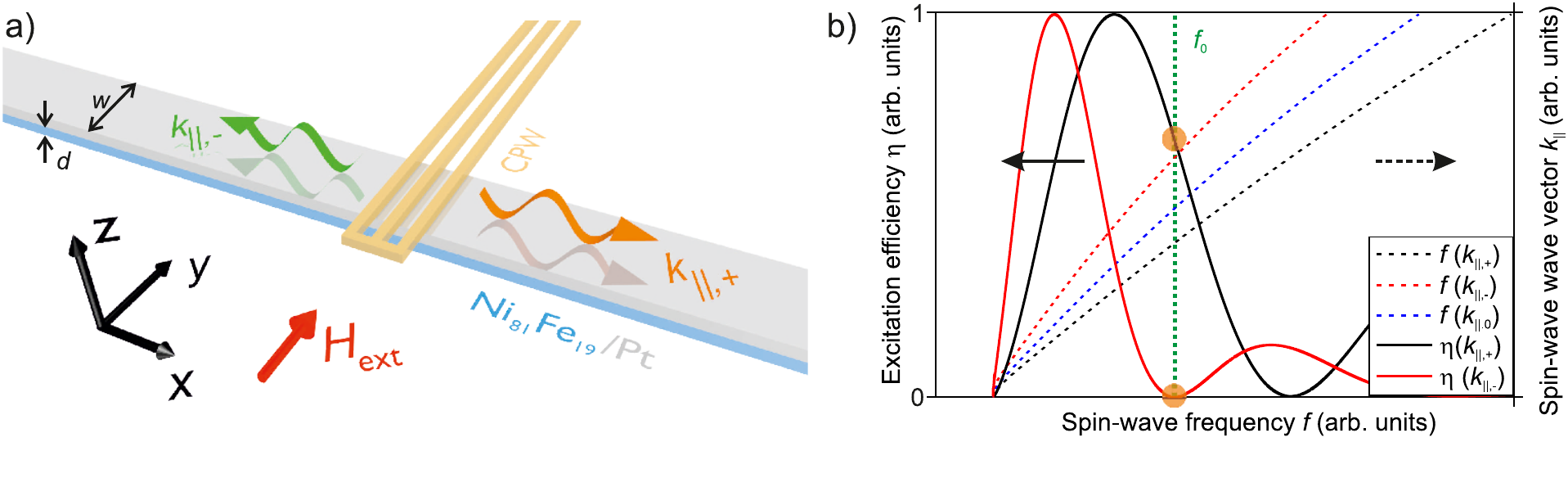}}
    \end{center}
	  \caption{\label{Fig1}(Color online) a) Schematic layout of the investigated transversely magnetized Ni$_{81}$Fe$_{19}$(Py)/Pt waveguide. The interfacial Dzyaloshinskii-Moriya interaction (iDMI) leads to a frequency splitting of waves running in the $\pm x$-direction with $k_{||,\pm}$. The coplanar waveguide (CPW) is introducing a wave-vector selective excitation. b) Schematic illustration of the concept of the uni-directional emission: The dashed lines show the spin-wave dispersion relation of the fundamental waveguide mode. The iDMI leads to a lifting of the degeneracy of the wave vector for waves running to the left ($-x$) and to the right ($+x$). For a fixed frequency $f_0$ (vertical dashed line), this leads to different excitation efficiencies $\eta$, for waves running in the two opposite direction. These are schematically shown by the solid lines. A proper choice of $f_0$ results in the suppression of one emission direction, which is highlighted by the orange circles.}
\end{figure*}
The standard description of the waveguide modes in transversely magnetized waveguides is based on the dispersion relation for thin films developed in Ref. \cite{Kalinikos-1986-Disp} and has already been successfully applied to describe a large variety of experiments\cite{Pirro-2011-1,Bayer-2006,Vogt-PR}. The waveguide, which is schematically shown in Fig. \ref{Fig1}~(a) features two quantization directions: One across its thickness $d$ ($z$-direction) and one across its width $w$ (short axis, $y$-direction). The quantization across the thickness leads to the formation of perpendicular standing spin-wave (PSSW) modes. Commonly, and in particular for the manifestation of iDMI, the spin-wave waveguides are made from thin magnetic films. 
In contrast to the fundamental thickness mode, which corresponds to a homogeneous magnetization across the thickness of the waveguide ($k_z = 0$), the higher PSSW modes feature large exchange energies. Thus, they are separated from the fundamental thickness mode by several tens of GHz. Moreover, these modes feature a negligible dipolar dispersion and, thus, for rather small wave vectors they are essentially not propagating along the waveguide\cite{Kalinikos-1986-Disp,Stancil}. Thus, the fundamental thickness mode is the only mode of interest in terms of magnonic applications. In addition, one should note that for the ultra-thin films, the non-reciprocity induced by asymmetric interface anisotropies can be neglected\cite{Gladii-2016}.

In the following, we discuss exclusively the case of transversely magnetized waveguides, where the transverse magnetization is enforced by an externally applied bias field. In this geometry (also referred to as the Damon-Eshbach geometry), the iDMI causes a non-reciprocity in the wave-vector component $k_{||}$ along the propagation direction along the wire. This results in a wave-vector dependent frequency splitting between the two propagation directions\cite{Moon-2013}. 

In the waveguide geometry, the geometric confinement across the short axis leads to the formation of waveguide modes. For a given waveguide mode $n$, this is taken into account by attributing a fixed wave-vector component $k_\perp(n) = n \pi/w_\mathrm{eff} $ across the effective width $w_\mathrm{eff}$ of the waveguide. The effective width is, hereby, determined by the geometrical width of the waveguide as well as the dipolar pinning on the waveguide edges. This results in an inhomogeneous effective field across the waveguide width\cite{Guslienko-eff-width,Mockel-2008-1}. The effective width and the corresponding demagnetization field $\mu_0 H_\mathrm{demag}$ in the waveguide can be approximated following the approach derived in Refs. \cite{Mockel-2008-1,Joseph-1965}, neglecting the small canting of the magnetization at the waveguide edges and assuming that the waveguide is homogeneously magnetized along its short axis ($y$-direction) by the external field. Under this assumption, the effective-field distribution $\mu_0 H_\mathrm{eff}(y) = \mu_0 H_\mathrm{ext} - \mu_0 H_\mathrm{demag}(y)$ in the transversely magnetized waveguide is given by
\begin{widetext}
\begin{align}
\label{Eq:Beff}
\begin{split}
\mu_0H_\mathrm{eff}(y) &= \mu_0H_\mathrm{ext} - \mu_0 M_\mathrm{s} N_{zz,0}(y) = \mu_0H_\mathrm{ext} - \frac{\mu_0 M_\mathrm{s}}{\pi}\left[\mathrm{arctan}\left(\frac{d}{2y + w}\right)-\mathrm{arctan}\left(\frac{d}{2y - w}\right)\right],
\end{split}
\end{align}
\end{widetext}
with $N_{zz,0}$ representing the demagnetization of the homogeneously magnetized waveguide. The effective width $w_\mathrm{eff}$ is taken as the separation between the points where the effective field inside the waveguide drops significantly. In this work we define the effective width as the part of the waveguide cross-section wherein the effective field is at least $85\,\%$ of the value in the waveguide center. The effective field used in the calculations is the average value of the effective field within this width.

Due to the rather large sizes of the considered waveguides\footnote{We consider waveguides of widths and lengths of a few $100\,\mathrm{nm}$, which is large in comparison to the distances relevant for the potential tilt of the magnetization due do iDMI.} and the comparably large ratio between waveguide length and waveguide width, we neglect any influence of the considered comparably weak iDMI on the effective width or the effective field due to a magnetization canting at the waveguide edges\cite{Cubukcu-PRB-2016} in the analytical model. With $w_\mathrm{eff}$ and $\mu_0 H_\mathrm{eff}$, the dispersion is then calculated as a function of the wave-vector component $k_{||}$ and the angle $\theta_k$ between the spin-wave wave vector in the waveguide plane and the magnetization direction. In the transversely magnetized scenario, this angle is given by
\begin{align}
\label{Eq:Thetak}
\theta_k &= \mathrm{atan}(k_{||}/k_\perp(n)).
\end{align}
Consequently, in a ferromagnetic waveguide (in the absence of material-related magnetic anisotropies), the dispersion relation $f(\mathbf{k}$) of the $n$-th waveguide mode for $|\mathbf{k}|d < 1$ can be calculated\cite{Kalinikos-1986-Disp,Di-2015}:
\begin{align}
\begin{split}
\label{Eq:wk}
f_\pm(\mathbf{k}) &= \sqrt{(\omega_0+\omega_\mathrm{M}\lambda_\mathrm{ex}\mathbf{k}^2)(\omega_0+\omega_\mathrm{M}\lambda_\mathrm{ex}\mathbf{k}^2+\omega_\mathrm{M}F_{00})}\\
&\pm\frac{2\gamma}{M_\mathrm{s}}D_\mathrm{int}|\mathbf{k}|\mathrm{sin}(\theta_k),
\end{split}
\end{align}
with the wave vector $\mathbf{k} = (k_{||},k_\perp,0)$, the gyromagnetic ratio $\gamma$ in units of $\mathrm{Hz}\,\mathrm{T}^{-1}$, the vacuum permeability $\mu_0$, $\omega_0 = \gamma \mu_0 H_\mathrm{eff}$, $\omega_\mathrm{M} = \gamma \mu_0 M_\mathrm{s}$, where $M_\mathrm{s}$ is the saturation magnetization, and $\lambda_\mathrm{ex} = 2A_\mathrm{ex}/(\mu_0M_\mathrm{s}^2)$, where $A_\mathrm{ex}$ is the exchange constant. $F_{00}$ is given by
\begin{align}
\label{Eq:F00}
F_{00} = 1 + g(\mathbf{k})(\mathrm{sin}(\theta_k)^2-1)+\frac{\omega_\mathrm{M}g(\mathbf{k})(1-g(\mathbf{k}))\mathrm{sin}(\theta_k)^2}{\omega_0+\omega_\mathrm{M}\lambda_\mathrm{ex}\mathbf{k}^2}
\end{align}
with the function $g(\mathbf{k}) = 1-(1-\mathrm{exp}(|\mathbf{k}|d))/(|\mathbf{k}|d)$. The last part in Eq. \ref{Eq:wk} is the implementation of the iDMI discussed in, for instance, Ref. \cite{Belmeguenai-PRB-2015,Di-2015}, where $D_\mathrm{int}$ is the effective iDMI strength in units of $\mathrm{J}\,\mathrm{m}^{-2}$. The latter is the effective volume strength of the iDMI, which is, for instance, originating from the Py/Pt interface. Thus, it depends on the used material combination, the exact interface and the thickness of the ferromagnetic layer. Moreover, it should be noted that, since the iDMI is a vectorial property, the sign of $D_\mathrm{int}$ depends on the orientation of the DMI-vector $\mathbf{D}$ and, consequently, on the orientation of the interface normal $\mathbf{\hat{n}}$\cite{Moon-2013}. Hence, the sign of $D_\mathrm{int}$ depends, for instance, on the stacking order of the layer system (e.g., it is opposite in Py/Pt and Pt/Py).

In Eq. \ref{Eq:wk}, the $\pm$-sign distinguishes the propagation to the right ($\mathbf{k}\cdot\mathbf{\hat{x}} > 0$) from the propagation to the left ($\mathbf{k}\cdot\mathbf{\hat{x}} < 0$) for a positive applied bias field ($\mathbf{M}$ aligned along the $y$-direction). If the field and, consequently, the magnetization are rotated by $180^\circ$, the two directions change their signs in Eq. \ref{Eq:wk} (i.e. $\mp$ instead of $\pm$). It should be noted that in this geometry, $|\mathbf{k}|\mathrm{sin}(\theta_k) = k_{||}$. Hence, the DMI leads to a splitting of the spin-wave frequency $\Delta f$ given by
\begin{align}
\label{Eq:Df}
\Delta f = f_+ - f_- = \frac{4\gamma}{M_\mathrm{s}}D_\mathrm{int}k_{||}.
\end{align}

From the spin-wave dispersion relation, the spin-wave group velocity $v_\mathrm{g}$ can be obtained straightforward by the differentiation of $f(\mathbf{k})$ with respect to $k_{||}$
\begin{align}
\label{Eq:vg}
v_{\mathrm{g},\pm} = \frac{\partial \omega_\pm(\mathbf{k})}{\partial k_\mathrm{||}} = 2\pi\frac{\partial f_\pm(\mathbf{k})}{\partial k_\mathrm{||}},
\end{align}
which we performed numerically from the calculated spin-wave dispersion relations. It should be noted that the iDMI leads to a constant offset of the group velocity in the two propagation directions $\Delta v_\mathrm{g} = 4\gamma/M_\mathrm{s}\cdot D_\mathrm{int}$, independent of the spin-wave wave vector.

From the phenomenological loss theory\cite{Stancil}, the spin-wave lifetime $\tau_0$ in the absence of iDMI follows to
\begin{widetext}
\begin{align}
\begin{split}
\label{Eq:Tau0}
\tau_0 &= \left(\alpha \omega(\mathbf{k}) \frac{\partial \omega(\mathbf{k})}{\partial \omega_0} \right)^{-1} = \left(2\pi\alpha\left(\omega_0+\omega_\mathrm{M}\lambda_\mathrm{ex}\mathbf{k}^2+\frac{\omega_\mathrm{M}(1+g(\mathbf{k}))(\mathrm{sin}(\theta_k)^2-1)}{2}\right)\right)^{-1},
\end{split}
\end{align}
\end{widetext}
with the Gilbert damping parameter $\alpha$ and $\omega(\mathbf{k}) = 2\pi f(\mathbf{k})$. To incorporate the effect of iDMI on the spin-wave lifetime, we follow the approach presented in Ref. \cite{Di-2015}:
\begin{align}
\label{Eq:Taupm}
\tau_\pm = \frac{\tau_0}{1 \pm \frac{\Delta f}{2 f_0}},
\end{align}
where $f_0$ is the spin-wave frequency in the absence of iDMI, which is calculated from Eq.~\ref{Eq:wk} with $D_\mathrm{int} =0$. With the spin-wave lifetime and group velocity, the exponential spin-wave amplitude decay length $\delta_\pm = v_\mathrm{g,\pm} \cdot \tau_\pm$ can be computed.

Another important parameter of the waveguide modes in the spin-wave waveguides is their ellipticity, which, amongst others, influences the threshold for the parametric instability\cite{LVov,Braecher-2011-1,Schloemann-1960-1,Kostylev-1995-1} and which is given by\cite{Melkov-1996-1}
\begin{widetext}
\begin{align}
\begin{split}
\label{Eq:Ell}
\epsilon &= 1-\frac{|m_z|^2}{|m_x|^2} =\frac{\omega_\mathrm{M} |g(\mathbf{k})\mathrm{sin}(\theta_k)^2-(1-g(\mathbf{k}))|}{\omega_0+\omega_\mathrm{M}\lambda_\mathrm{ex}\mathbf{k}^2+\frac{\omega_\mathrm{M}}{2}(g(\mathbf{k})\mathrm{sin}(\theta_k)+1-g(\mathbf{k}))+\frac{\omega_\mathrm{M}}{2}|g(\mathbf{k})\mathrm{sin}(\theta_k)^2-(1-g(\mathbf{k}))|},
\end{split}
\end{align}
\end{widetext}
with the components $m_x$ and $m_z$ of the dynamic magnetization. Implicitly, the ellipticity becomes different for waves running in the two directions along the waveguide due to the fact that, for a fixed frequency $f$, $k_{||,+} \neq k_{||,-}$.

In addition to the spin-wave frequency and propagation characteristics, we analyze the spin-wave excitation in the presence of iDMI. For this, we follow the approach presented in Refs. \cite{Mockelantenne,Pirro-2011-1} and incorporate the iDMI-dependent changes of the spin-wave dispersion into the calculations. We aim at a description of the excitation efficiency, a relative measure of how efficiently the antenna can exert a net torque on the static magnetization at a given frequency and wave-vector combination. This efficiency is linked to the magnetic susceptibility and the spatial extent of the excitation source\cite{Kostyant}. We assume an excitation source which is homogeneous across the width of the approximately homogeneously magnetized waveguide. Consequently, the excitation efficiency, for an odd waveguide mode $n$ at a given frequency is given by
\begin{widetext}
\begin{align}
\label{Eq:Exeff}
\begin{split}
\eta_\pm(k_{||,\pm}) &= \left|\frac{1}{n} |b_k| \left(\frac{f(k_{||,\pm})}{\gamma}\mp\frac{1}{\mu_0 M_\mathrm{s}}\left(\mu_0 H_\mathrm{eff}^2-\frac{f(k_{||,\pm})^2}{\gamma^2}\right)\right)\right|,
\end{split}
\end{align}
\end{widetext}
and it vanishes for even modes, since they don't exhibit a net magnetic moment integrated across the width of the waveguide. Here, $|b_k|$ represents the wave-vector spectrum of the effective field created by the excitation source, which is assumed to be the Oersted field created by the current flowing through the CPW\cite{Chumakov}. It is obtained via the Fourier transformation of the spatial distribution of the Oersted field $b_\mathrm{CPW}$(see Appendix for further details). The factor $1/n$ results from the decreasing total magnetic moment of the (uneven) waveguide modes. Due to the iDMI-induced wave-vector dependent frequency splitting, the excitation for waves running to the left and waves running to the right features a different periodicity in the spin-wave frequency $f$. This is the basic concept of the uni-directional spin-wave emission discussed in this article, which is schematically illustrated in Fig.~\ref{Fig1}~(b).

With all these ingredients, we are able to present the main equation of this work: The expected spin-wave amplitude, proportional to the dynamic component $m_x$, at a given distance $x$ from the excitation source located at $x_0$:
\begin{align}
\label{Eq:Examp}
m_x(f_\pm)  \propto\sum_n^{n_\mathrm{max}} \eta_\pm(f_\pm,n) \mathrm{exp}\left({\frac{x-x_0}{\delta_\pm(f_\pm,n)}}\right).
\end{align}
Here, the sum is over the various waveguide modes $n$ contributing to the overall amplitude. In the case of an elliptic magnetization trajectory, the out-of-plane component $m_z$ is related to $m_x$ via Eq. \ref{Eq:Ell} and a phase shift of $\pi/2$. The emitted intensity is given by the square of the total amplitude $|m_x|^2=|m_z|^2/(1-\epsilon)$. This equation constitutes the main result of the analytical model in this work, since it allows to predict the relative spin-wave amplitude within the waveguide, including the iDMI-dependent changes of the spin-wave excitation and propagation characteristics. To evaluate the intensity measured by a detector, such as an additional antenna, the intensity has to be multiplied with the detection function of the detector. For a microstrip detector, this is again given by the square of the Fourier transform of its field distribution and another factor $n^{-2}$ accounting for the reduced effective moment of the waveguide modes. Hence, for a homogeneous detector sensitive to the spin-wave intensity, the influence of higher waveguide modes falls off with $n^{-4}$. Thus, in practice, most characteristics are determined by the fundamental waveguide mode $n = 1$. 

It should be noted that the presented analytical theory does not account for the presence of edge modes in the waveguides and does not account for a potential dipolar interaction between the waveguide modes, such as avoided crossings\cite{Kalinikos-1986-Disp}. For a precise description of all peculiarities of the spin-wave propagation in a waveguide, these effects should be taken into account. However, they involve a more elaborate derivation of the spin-wave dispersion\cite{Kostylev,Grigoryeva,Micromag} which can generally not be performed analytically anymore. The presented theory has proven to be a very versatile tool for the description of most phenomena observed in spin-wave waveguides in the absence of iDMI unless $k_{||} \rightarrow 0$. As we will show in the following, also in the presence of iDMI, in a wide wave-vector range, the derived analytical theory is in good agreement with the numerical simulations, which take the interaction between the modes as well as the inhomogeneous magnetization distribution into account. 

\section{Numerical simulations}
\label{Num}

We compare the aforementioned analytical calculations to micromagnetic simulations using MuMax3\cite{MuMax-2014-1}. The iDMI is implemented in this micromagnetic package by an effective field\cite{MuMax-2014-1,Bogdanov-2001}
\begin{align}
\label{Eq:BDMI}
\mathbf{B}_\mathrm{DMI} = \frac{2D_\mathrm{int}}{M_\mathrm{s}} \left(\frac{\partial m_z}{\partial x}, \frac{\partial m_z}{\partial y}, -\frac{\partial m_x}{\partial x}-\frac{\partial m_y}{\partial y}\right),
\end{align}
with the application of boundary conditions based on Ref. \cite{Rohart-PRB-2013}. 

The simulated geometry, which is also used in the analytical calculations, consists of a transversely magnetized waveguide made from Ni$_{81}$Fe$_{19}$(Py)/Pt. We assume a geometrical width of the waveguide of $w = 600\,\mathrm{nm}$ and a Py thickness of $d = 2.5\,\mathrm{nm}$. The Pt is implemented by the creation of an iDMI with the strength $D_\mathrm{int} = 0.175\,\mathrm{mJ}\,\mathrm{m}^{-2}$. This value is in line with recent experimental findings in Py/Pt layer systems of comparable thickness\cite{Nembach-2015,Stashkevich-2015}. Moreover, we account for the spin-pumping induced enhancement of the damping parameter in Py/Pt, which we assume to result in $\alpha = 0.02$. For the simulations, a $30\,\mu\mathrm{m}$ long waveguide is discretized in $8192\times128$ cells in the $x$-$y$-plane (i.e., cell size of about $3.7\times4.7\,\mathrm{nm}^2$). One cell of $2.5\,\mathrm{nm}$ is assumed along the $z$-direction, representing the uniform magnetization across the layer thickness. Reference simulations have been performed in an identical waveguide without iDMI. In all calculations and simulations presented here, a magnetic field of $\mu_0 H_\mathrm{ext} = 25\,\mathrm{mT}$ applied in the $+y$-direction has been assumed. A change of the field-direction interchanges the asymmetry in the antenna excitation as well as the influence of the iDMI with respect to $\pm k_\mathrm{||}$.

Two different types of simulations of the magnetization dynamics in the waveguide have been performed. In the first type, the magnetization is excited by a short Gaussian pulse (width $\Delta t = 20\,\mathrm{ps}$), which is applied locally in a rectangle with dimension $10\times600\,\mathrm{nm}^2$ in the center of the waveguide. This short and highly localized pulse gives access to a wide wave-vector and frequency excitation-span, which is used to determine the spin-wave dispersion and the related parameters in the waveguide. The space-resolved temporal evolution of the magnetization dynamics is simulated for a total of $20\,\mathrm{ns}$, the pulse being applied centered around $1\,\mathrm{ns}$ of simulation time. Snapshots of the space-resolved magnetization in the waveguide are stored every $20\,\mathrm{ps}$. Since, as mentioned above, the excitation and detection of higher modes in practical geometries falls off with $n^{-2}/n^{-4}$ (amplitudes/intensities) and because the higher waveguide modes feature smaller group velocities, we focus our attention on the fundamental waveguide mode $n = 1$, which is the most interesting mode for practical applications due to these facts. To do so, we integrate the obtained $x$-$y$-maps of the dynamic magnetization components across the $y$-direction, thus, reducing the influence of higher modes $n > 1$ to the recorded dynamics significantly. Subsequently, we perform fast Fourier transformations (FFT) of the data in time and space to obtain the magnetization dynamics in the frequency-space- and frequency-wave-vector domain.

From the frequency-wave-vector domain, the spin-wave dispersion relation $f_\pm(k_{||,\pm})$ of the fundamental mode is determined by extracting the frequency  with maximum Fourier amplitude at a given wave-vector $k_{||,\pm}$ in the two propagation directions. The spin-wave group velocity is computed from the derivative of the dispersion relation. The ellipticity is evaluated by determining the Fourier amplitude at a given frequency for the dynamical $m_z$ and $m_x$ components and by evaluating Eq. \ref{Eq:Ell} for both propagation directions. The spin-wave decay length is analyzed in the frequency-space domain by fitting an exponential decay function to the spatial intensity distribution for a given frequency and directly extracting the decay lengths $\delta_\pm$.

In a second set of simulations, we mimic the spin-wave excitation by means of a short-circuited coplanar waveguide (CPW), which is schematically depicted in Fig. \ref{Fig1}. The simulated CPW features a width and a thickness of the conducting wires of $50\,\mathrm{nm}$ each, which are spaced by $300\,\mathrm{nm}$. The Oersted field components $h_x$ and $h_z$ created by the CPW are calculated at the bottom of the CPW (i.e., in direct contact with the Py/Pt) from the (static) Ampere's law, assuming that the outer strips (ground lines) are out of phase with the central wire (signal line) (see Appendix for further detail). The same field distribution is used to calculate the Fourier spectrum $b_k$ and the direction-dependent excitation spectrum according to Eq. \ref{Eq:Exeff}. In the simulation, the magnetization is excited by $10\,\mathrm{ns}$ long pulses with carrier frequencies ranging from $4\,\mathrm{GHz}$ to $20\,\mathrm{GHz}$ in steps of $250\,\mathrm{MHz}$. The pulses are modeled by
\begin{align}
\label{Eq:Pulse}
\begin{split}
\tilde{h}_{x,z} &= 0.5h_{x,z}(x)(\mathrm{tanh}(2f_\mathrm{rise}(t-t_0))\\
&\cdot \mathrm{tanh}(2f_\mathrm{rise}(t_0+\Delta t-t))+1)\cdot\mathrm{sin}(2\pi ft)
\end{split}
\end{align}
with a rising frequency of $f_\mathrm{rise} = 5\,\mathrm{GHz}$, a duration $\Delta t = 10\,\mathrm{ns}$ and $t_0 = 1\,\mathrm{ns}$. The spin-wave dynamics are saved every $20\,\mathrm{ps}$ spatially averaged over regions of $10\times600\,\mathrm{nm}^2$ situated in distances of $\pm j \cdot 250\mathrm{nm}$ from the antenna with $j$ being an integer. The data are analyzed by analyzing the maximum amplitude in the time domain as well as the integrated amplitude of the FFT for each of this regions in space. In this analysis, the first and last nanosecond of the pulse are omitted to eliminate artifacts originating from the rising and falling edges of the pulse. The detector-regions with a length of $10\,\mathrm{nm}$ in the $x$-direction do not falsify the recorded wave-vector spectrum notably since their size is much smaller than the smallest excited wavelength in the presented simulations. Nevertheless, the averaging over the width of the wire in the detector regions leads again to a suppression of the influence of the higher waveguide modes by an additional factor of $n^{-1}$ in amplitude due to the decreased detection efficiency.

\section{Comparison of numerical and analytical results}
\label{Res}

\begin{figure}[t]
	  \begin{center}
    \scalebox{1}{\includegraphics[width=0.5\linewidth, clip]{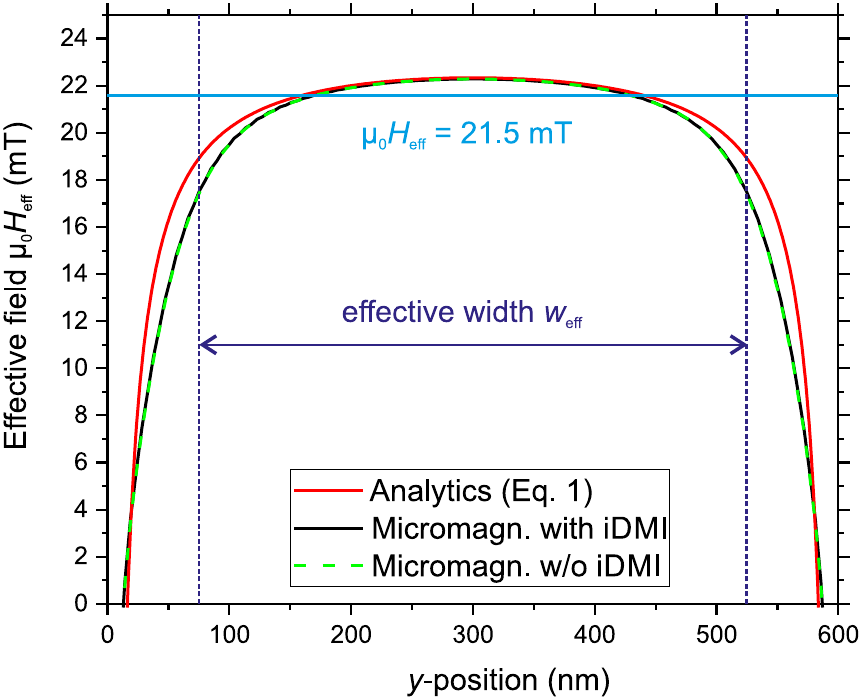}}
    \end{center}
	  \caption{\label{Fig2}(Color online)Effective field $\mu_0H_\mathrm{ext}$ in the $y$-direction as a function of the position across the $600\,\mathrm{nm}$ wide waveguide according to Eq. \ref{Eq:Beff} (red line) and according to the micromagnetic simulations (black and dashed green lines).  The horizontal magenta line represents the average effective field $\mu_0H_\mathrm{eff} = 21.5\,\mathrm{mT}$ in the central region of the waveguide, which is enclosed by the effective width of the waveguide.}
\end{figure}
In the following Section, we compare the analytically calculated spin-wave properties with the results from our numerical simulations. For the simulations as well as the calculations we use the following material parameters for the Py: Saturation magnetization $M_\mathrm{s} = 800\,\mathrm{kA}\,\mathrm{m}^{-1}$, exchange constant $A_\mathrm{ex} = 13\,\mathrm{pJ}\,\mathrm{m}$ and, as mentioned above, a Gilbert damping parameter $\alpha = 0.02$ as well as an effective iDMI constant of $D_\mathrm{int} = 0.175\,\mathrm{mJ}\,\mathrm{m}^{-2}$. We start by comparing the effective field as a function of the position across the waveguide derived from Eq. \ref{Eq:Beff} with the result of the numerical simulations. The respective effective fields in the $y$-direction are shown in Fig.~\ref{Fig2}. As can be seen from the figure, the field distributions are qualitatively fairly similar. Towards the edges, the profile obtained from the micromagnetic simulations shows a smoother reduction of the effective field which extends slightly further into the center of the waveguide. This is due to the fact that the micromagnetic simulations take into account that the magnetization is changing its orientation at the edges to minimize the dipolar stray fields, an effect which is not taken into account in Eq. \ref{Eq:Beff}. Nevertheless, the two approaches lead to quantitatively the same result in the central region of the waveguide, justifying the use of the analytical formalism to derive the effective magnetic field $\mu_0 H_\mathrm{eff} = \mu_0 H_\mathrm{ext} - \mu_0 H_\mathrm{demag}$ in the waveguide. As mentioned above, the values of the effective field and the effective width of the waveguide used in the analytical model are associated with the waveguide region where the effective field measures $85\,\%$ of its maximum value in the center. The borders of this region give the effective width, which is $w_\mathrm{eff} \approx 450\,\mathrm{nm}$ in the present case. The effective field value of $\mu_0 H_\mathrm{eff} = 21.5\,\mathrm{mT}$ is obtained by the spatial average of the effective field over this central region. Please note that, in agreement with the calculations in Ref.~\cite{Sanchez-2014}, the assumed weak iDMI does not impose significant changes on the magnetization distribution within the waveguide, which reflects in the fact that the effective fields within the waveguide with and without iDMI are basically the same (cf. black solid and green dashed curves in Fig.~\ref{Fig2}).

\begin{figure}[t!]
	  \begin{center}
    \scalebox{1}{\includegraphics[width=0.9\linewidth, clip]{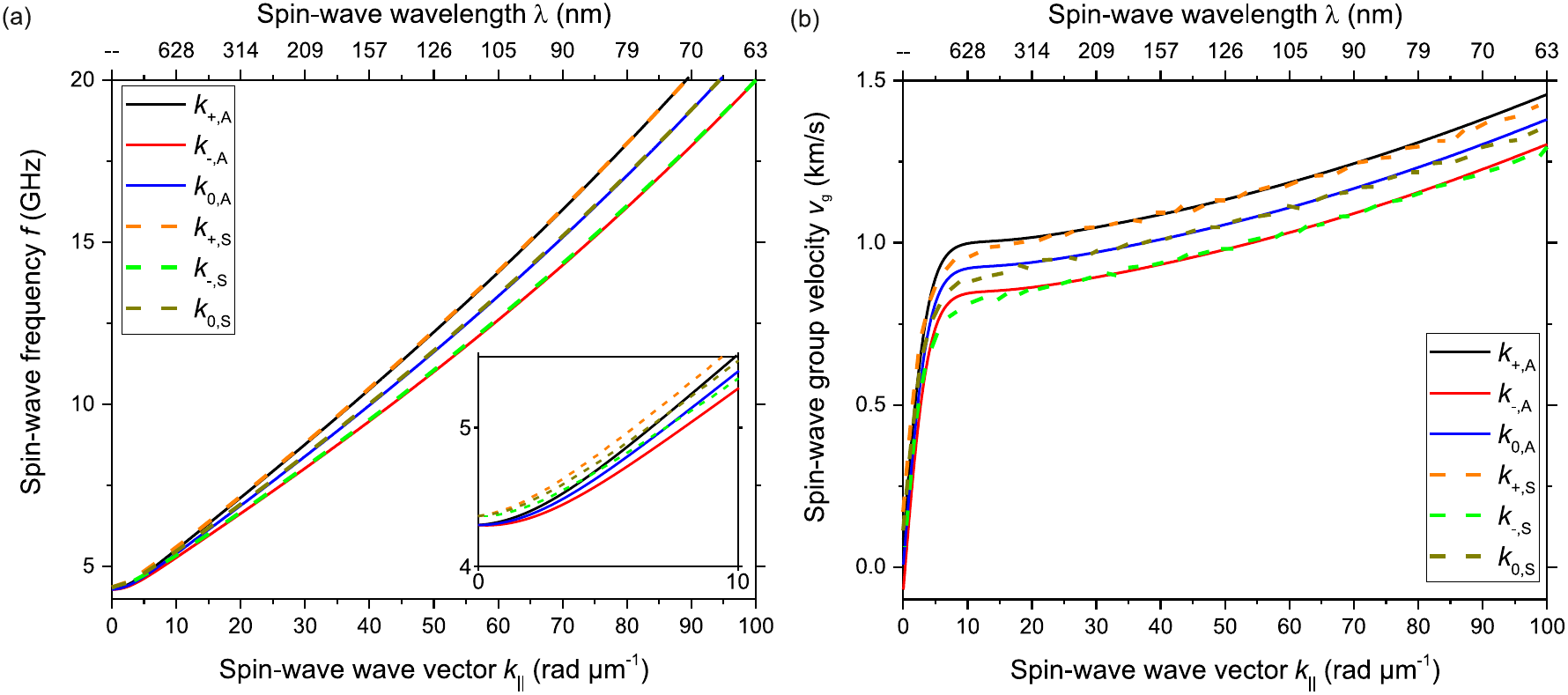}}
    \end{center}
	  \caption{\label{Fig3}(Color online) a) Spin-wave dispersion relations for the two emission directions with iDMI and in the absence of iDMI. Solid lines are the result of the analytical calculations (abbreviation 'A'), dashed lines are simulated values (abbreviation 'S'). The index '$\pm$' refers to calculations/simulations with iDMI for waves propagating in the $\pm x$-direction and '0' to its absence. The inset shows the discrepancy at small wave-vectors which mainly stems from a underestimation of the frequency origin by about $60\,\mathrm{MHz}$ in the analytical calculations. b) Corresponding group velocities.}
\end{figure}
The basic parameters for the analytical calculations being determined this way, we now move on to a comparison of the dispersion relation of the fundamental mode as well as its group velocity. The analytically and numerically derived dispersion relations are shown in Fig.~\ref{Fig3}~(a). As can be seen from the figure, the dispersion relations are in excellent agreement in a wide wave-vector range and only for very small wave vectors, there is a visible deviation. The latter is connected to the approximation of the effective width. This has a notable influence for waves with small wave-vectors $k_{||}$ along the wire ($\theta_k$ close to 0, i.e., $k_{\perp} \gg k_{||}$) and, in particular, for the origin of the spin-wave dispersion relation which is underestimated in the analytical calculations by about $60\,\mathrm{MHz}$ (see inset in Fig.~\ref{Fig3}). In this wave-vector range, the numerical simulations are likely to be closer to reality, since they take the full internal field landscape as well as the non-uniform orientation of the magnetization into account. Nevertheless, for $k_{||} \gtrsim 15\,\mathrm{rad}\,\mu\mathrm{m}^{-1}$ the analytical description is adequate. Due to the small thickness, it also remains adequate for substantially larger $k_{||}$. This is due to the fact that $|\mathbf{k}|\cdot d$ remains significantly smaller than $1$ in the investigated wave-vector range. Consequently, the investigated wave vectors stay far below the limit of the analytical formalism at $|\mathbf{k}|\cdot d\approx 1$\cite{Kalinikos-1986-Disp}. The agreement and discrepancy of the analytically and the numerically computed dispersion relations are reflected in the derived group velocities shown in Fig.~\ref{Fig3}~(b): For small $k_{||}$ there is a small deviation of the analytically calculated group velocities, which vanishes for $k_{||} \gtrsim 15\,\mathrm{rad}\,\mu\mathrm{m}^{-1}$. 

\begin{figure}[t!]
	  \begin{center}
    \scalebox{1}{\includegraphics[width=0.9\linewidth, clip]{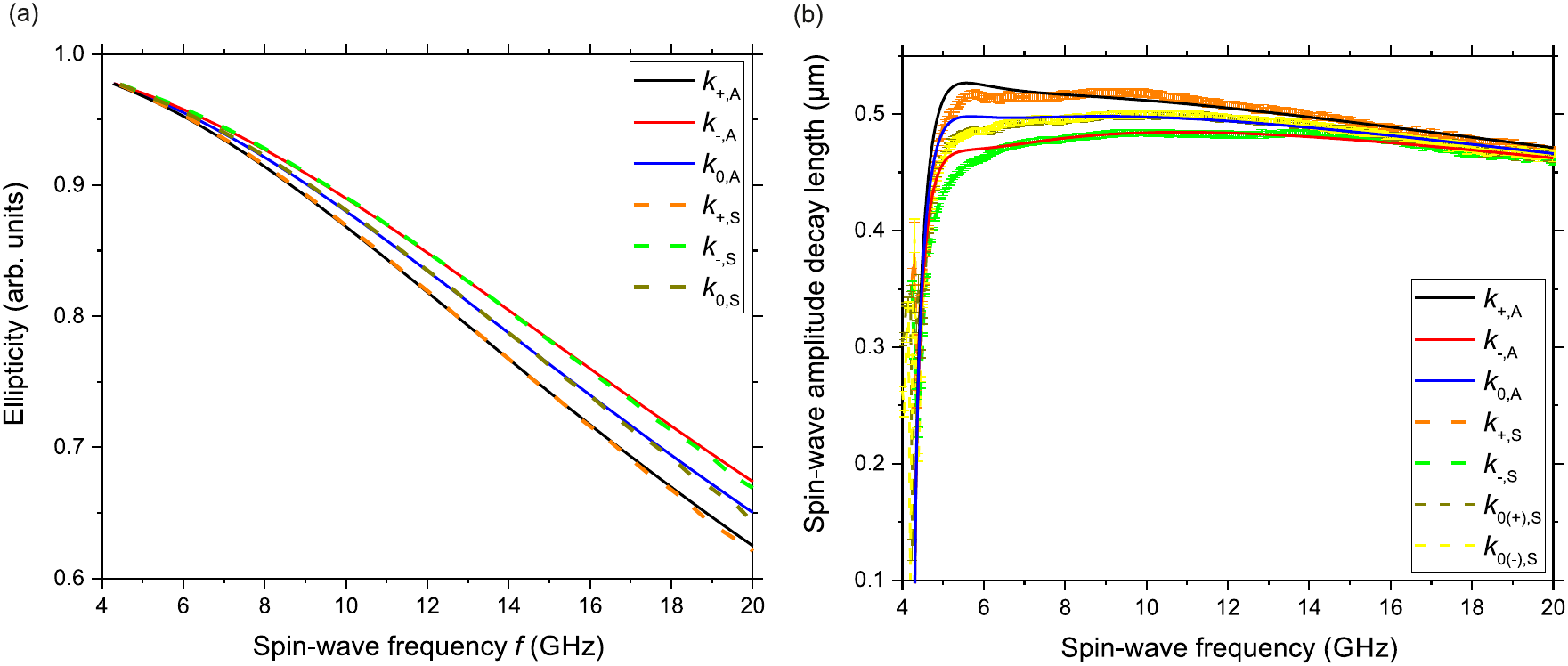}}
    \end{center}
	  \caption{\label{Fig4}(Color online) a) Ellipticity and b) spin-wave amplitude decay length of the fundamental waveguide mode as a function of frequency. Solid lines are the result of the analytical calculations (abbreviation 'A'), dashed lines are simulated values (abbreviation 'S'). The index '$\pm$' refers to calculations/simulations with iDMI for waves propagating in the $\pm x$-direction and '0' to its absence. The error bars in (b) are given by the uncertainty of the determination of $\delta$ from an exponential fit of the simulated data.}
\end{figure}

Next we address the ellipticity and the spatial decay of the fundamental waveguide mode. In Fig.~\ref{Fig4}~(a), we compare the analytical values of the ellipticity to the numerical values. As can be seen from the figure, the frequency-dependent ellipticity is in almost perfect agreement between the two, indicating that, in this case, the iDMI in the present geometry only results in the wave-vector dependent splitting and does not influence the ellipticity for a given wave vector by itself. The obtained spin-wave amplitude decay lengths are shown in Fig.~\ref{Fig4}~(b). Please note that in the figure, we show both emission directions in the absence of iDMI for completeness, proving that the two emission directions are indeed equal. Once again, we find a small disagreement between the numerical simulations and the analytical calculations for small frequencies. This is mainly caused by the difference in group velocity visible in Fig. \ref{Fig3}~(b). In contrast, for large frequencies, the decay lengths are in very good agreement. This implies that the calculation of the spin-wave lifetime via Eq. \ref{Eq:Taupm} is in good agreement with the spin-wave lifetime obtained in the numerical simulations. Hence, the analytical formalism above is well suited to predict also the spin-wave ellipticity and relaxation frequency, which are important parameters for the spin-wave propagation and the spin-wave amplification by parallel parametric amplification.

\section{Application to the design of uni-directional spin-wave emitters}
\label{Uni}

Ultimately, we want to use the analytical formalism and the numerical simulations to derive the spin-wave excitation by a CPW, highlighting the use of such a wave-vector selective element for the creation of uni-directional spin-wave emission in the presence of iDMI. As mentioned above, the CPW is assumed to have a wire-width of $50\,\mathrm{nm}$ with a center-to-center spacing of $300\,\mathrm{nm}$ between the wires. This leads to a periodic excitation spectrum with minima at $k_{\mathrm{min},l} = l\frac{2 \pi}{s}$ with $l$ being an integer. Due to the frequency-splitting for $k_\mathrm{||,\pm}$, this means that for one emission direction, the minima are situated at different frequencies than for the other direction. This way, an efficient emission in one direction can be achieved while the emission in the other direction is at its minimum. In a two antenna configuration for excitation and detection, this results in a diode-like behavior, where microwave signals can be transmitted in only one direction via spin waves. 

\begin{figure}[t!]
	  \begin{center}
    \scalebox{1}{\includegraphics[width=0.9\linewidth, clip]{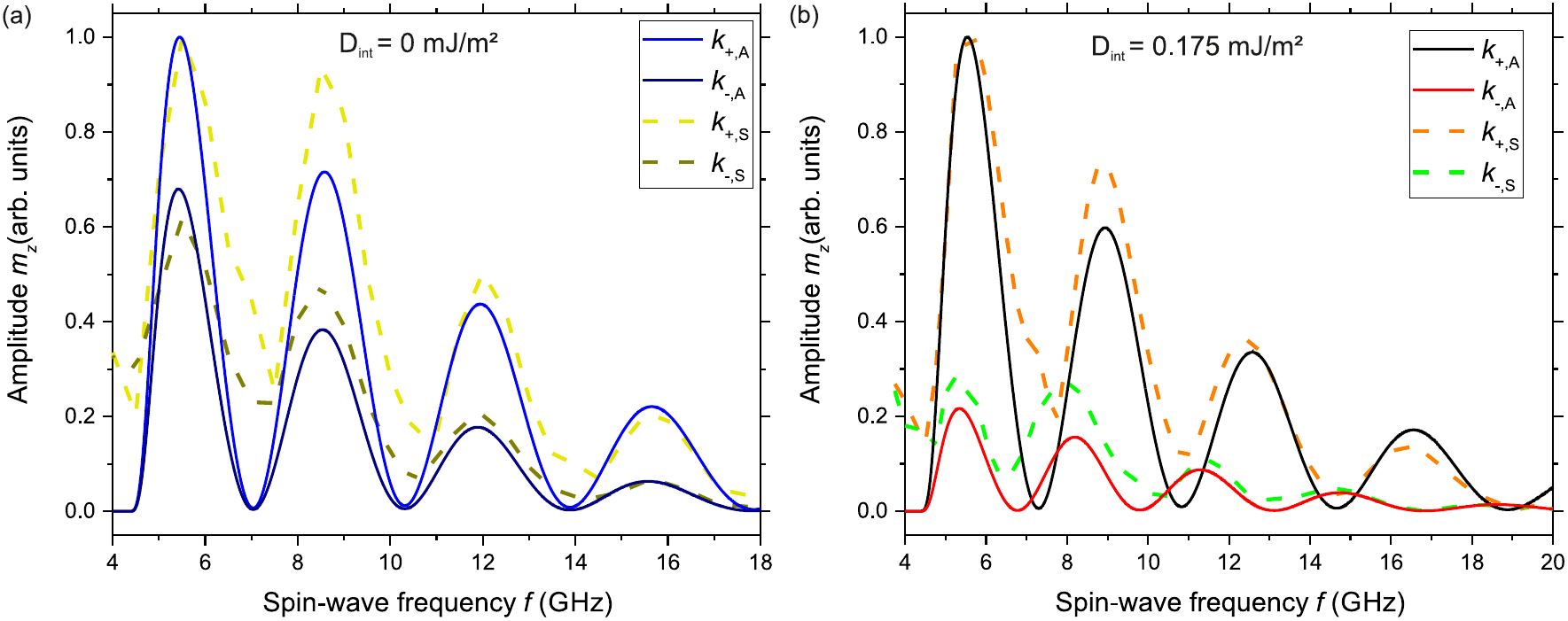}}
    \end{center}
	  \caption{\label{Fig5}(Color online) Spin-wave amplitude spectra at a distance of $4.5\,\mu\mathrm{m}$ from the center of the CPW a) in the absence of iDMI and b) in the presence of iDMI. Solid lines are the result of the analytical calculations (abbreviation 'A'), dashed lines are simulated values (abbreviation 'S'). The index '$\pm$' refers to calculations/simulations for waves propagating in the $\pm x$-direction.}
\end{figure}

The results of the simulated excitation spectra together with the analytical calculated ones are shown in Fig.~\ref{Fig5}. The simulated amplitude has been extracted from the dynamic $m_z$-component at a distance of $4.5\,\mu\mathrm{m}$ from the center of the CPW. The analytical calculations have been performed by evaluating Eq. \ref{Eq:Examp} taking into account the first 9 waveguide modes and with $x - x_0 = 4.5\,\mu\mathrm{m}$ and by accounting for the fact that the ellipticity results in a wave-vector dependent reduction of the $m_z$ component with respect to $m_x$\footnote{For this, the predictions of Eq. \ref{Eq:Examp} are multiplied with ($\sqrt{1-\epsilon} = |m_z|/|m_x|$ to be comparable to $m_z$. We chose a comparison with the dynamic $z$-component for convenience. A comparison with of the simulated $m_x$ component with Eq.~\ref{Eq:Examp} is also in good quantitative agreement.}. Figure \ref{Fig5}~(a) shows the results in the absence of iDMI. As can be seen from the figure, the periodicity of the CPW leads to the expected periodic excitation spectrum. The asymmetry in amplitude between the left and the right emission direction stems from the CPW excitation, namely from the interplay of the excitation by the out-of-plane and the in-plane Oersted field created by the antenna. This is reflected by the $\pm$ term in Eq. \ref{Eq:Exeff}. Nevertheless, in the absence of iDMI (or other effects lifting the frequency degeneracy between $k_{||,+}$ and $k_{||,-})$, the maxima and minima are always at the same wave vectors. The analytical calculations are in good agreement with the numerical simulations. The most pronounced deviation lies in the fact that the minima in the analytic calculations are much sharper than the ones observed in the simulation. This is a consequence of the finite linewidth of the spin waves. While this is not included in the analytic formalism stated above, this is taken into account in the micromagnetic simulation and leads to a broadening of the excitation spectra. Consequently, the minima are not as sharp as predicted analytically.

In Fig.~\ref{Fig5}~(b), iDMI is included into the calculations/simulations. As expected, this results in a displacement of the minima in one emission direction in frequency with respect to the other, a behavior well visible in the calculated as well as the simulated curve. As a consequence, the minimum of one direction can coincide with a significant emission in the other. This is, for instance, the case around $f \approx 13\,\mathrm{GHz}$, where the emission in the $k_{||,-}$ direction is minimal while the emission in the $k_{||,+}$ direction is maximal. Amplitude ratios exceeding a factor of 10 can easily be achieved, which can be further enhanced by employing a second, identical CPW for detection to obtain the same suppression in the detection efficiency. With this, intensity-asymmetries larger than $10^4$ are easily realized and a very efficient uni-directionality can be achieved. Hereby, the effective emission direction can be interchanged by changing the sign of the externally applied bias field. On the other hand, by changing, for instance, the operation frequency, a basically symmetric emission can be realized, as is the case for $f \approx 7.75\,\mathrm{GHz}$ in Fig.~\ref{Fig5}~(b). By comparing Figs.~\ref{Fig5}~(a) and (b), it is evident that in the presented situation, the iDMI enhances the asymmetry in the amplitude of the emission, independent of the horizontal shift along the frequency axis. Consequently, it leads to an overall enhancement of the unidirectional emission by the CPW in a large wave-vector range. This is due to the fact that the iDMI-induced asymmetry in the propagation characteristics (cf. Fig.~\ref{Fig4}~(b)) and the asymmetric CPW excitation add up. In contrast, if the sign of the iDMI is inverted, for instance by replacing the Py/Pt by a Pt/Py layer, the effects oppose and an overall more symmetric (in terms of maximal amplitude) emission can be achieved\footnote{A rotation of the field changes the efficient direction for both, the excitation asymmetry and the iDMI asymmetry and, hence, simply interchanges the efficient direction. In contrast, if the sign of the iDMI vector changes or the CPW is patterned on the opposite side of the waveguide ($z$-axis), the effects of the nonreciprocity of the antenna excitation and the asymmetric propagation characteristics due to the iDMI are opposing each other. In other words, whether the effects subtract or enhance is determined by the orientation of the iDMI-vector with respect to the placement of the excitation source on top or below the magnetic structure.}. Finally, we would like to remark that, again, the numerical simulation and the analytical calculation are in reasonable agreement, the main differences again originating from the effects of finite linewidth included into the simulations.

\section{Discussion}
\label{Disc}

In this last Section we would like to sum up the results and discuss their significance. As we have demonstrated in the previous Section, the simple analytical formalism presented above is well able to predict most of the features observed in the more realistic micromagnetic simulations. The main discrepancies arise from an imperfect analytical description for very small wave vectors and the influence of finite linewidth. For a real application, the first point does not really impose a major problem, since the miniaturization on the one hand and the wave-vector dependence of the iDMI on the other hand suggest the use of rather large wave vectors. In this wave-vector range, the analytical model is accurate. The effects of finite linewidth hamper the possibility to predict the exact strength of the uni-directionality. However, they do not affect the possibility to optimize the emission characteristics: The parameters of the system (such as thickness of the magnetic material, geometry of the antenna, geometric width of the waveguide, sign and, with some limitations, strength of the iDMI, material parameters of the magnetic material etc.) as well as the parameters of the experiment (applied field and frequency) can be applied to the analytical calculations and the system can be tuned easily to shift the maximum at the desired operation frequency in one emission direction to the minimum of the other.

The presented analytical model allows for such a good prediction of the spin-wave properties because the considered iDMI is rather small. Consequently, its main impacts are the lifting of the frequency-degeneracy and the small changes of the spin-wave lifetime, which can well be described by the first order expansion which is the basis of Eq. \ref{Eq:Taupm}. Any influence of the iDMI on the magnetization configuration in the waveguide or the ellipticity of the magnetization trajectory itself can be neglected in the presented case. It should be noted that other mechanisms which lift the degeneracy of the spin waves in transversely magnetized waveguides, such as the influence of non-uniform surface anisotropy, can easily be implemented into the presented analytical calculations to extend the model to a prediction of their influence as well.

With the used model system of Py/Pt featuring a small iDMI, we have demonstrated that an efficient uni-directional spin-wave emission can be achieved by combining a wave-vector selective excitation with the iDMI-induced non-reciprocity of the spin-wave dispersion. By matching the feature sizes of the excitation source and the material parameters, the uni-directional emission can be tuned to the desired wave-vector range. In particular, it allows for the realization of such emitters even for dipolar-exchange waves and without the need for an involved texture in the spin-wave waveguide. The concept is also applicable to layer systems featuring a frequency non-reciprocity due to a different origin, such as a non-uniform surface anisotropy. We would like to point out that we choose Py/Pt as model system due to its large use in many experiments of spin waves in microstructures. We admit that in a real system of Py/Pt of this thickness, the damping of the waves could even be higher\cite{Ruiz-2015}, depending on the quality and the exact nature of the interface between the materials. Hence, for a real application, this material system is likely not the best choice due to the poor spin-wave propagation characteristics. Other material systems, such as CoFeB/Pt, which are known to exhibit a similar iDMI-strength and a larger spin-wave lifetime\cite{Ruiz-2015,Di-2015-APL, Tacchi-2016, Conca-2014}, and presently unknown or unexplored material combinations of low-damping materials such as the half-metallic Heusler compound CoMnFeSi\cite{Tomseb-2012, Tomseb-2015} with heavy metals could provide more practical systems which allow for a more efficient spin-wave propagation than Py/Pt. Since even with the weak iDMI presented in Py/Pt an efficient uni-directional emission can be achieved, the use of a system with larger iDMI-strength would allow for the use of thicker films, featuring higher group velocities and a lower influence of interfacial damping. The presented analytical formalism can be applied to these systems in a straightforward manner, allowing for the design of the optimum structure for a non-reciprocal, or, if desired, reciprocal spin-wave emission by a periodic excitation source of any kind. 

\section{Conclusions}
\label{Conc}

To conclude, we have presented an analytical formalism to predict the spin-wave properties and the spin-wave excitation by periodic excitation sources in micro-structured, transversely magnetized spin-wave waveguides in the presence of iDMI. We have compared the formalism to numerical simulations, demonstrating that the analytical calculations can readily predict the spin-wave system with a high accuracy. This way, the formalism allows for the straightforward design of uni-directional and bi-directional spin-wave emitters on the micro-scale and can be used readily to interpret experiments. We have demonstrated that the iDMI in asymmetric layer systems enables a directed spin-wave emission. This permits a guiding of the excited spin waves in a magnonic network in the desired direction, optimizing the energy flow and avoiding undesired spin-wave beams. The periodic effective field can also be provided by any other excitation source, such as the use of electrical fields for the spin-wave excitation. This allows for a scalable realization of such emitters in any desired wave-vector range compatible with state-of-the art patterning techniques and detection schemes.

\begin{acknowledgements}
Financial support by the spOt project (318144) of the EC under the Seventh Framework Programme, by the DFG (SFB/TRR 173 Spin+X) and by the Nachwuchsring of the TU Kaiserslautern is gratefully acknowledged.
\end{acknowledgements}

\appendix*

\section*{Appendix}

\begin{figure}
	  \begin{center}
    \scalebox{1}{\includegraphics[width=0.5\linewidth, clip]{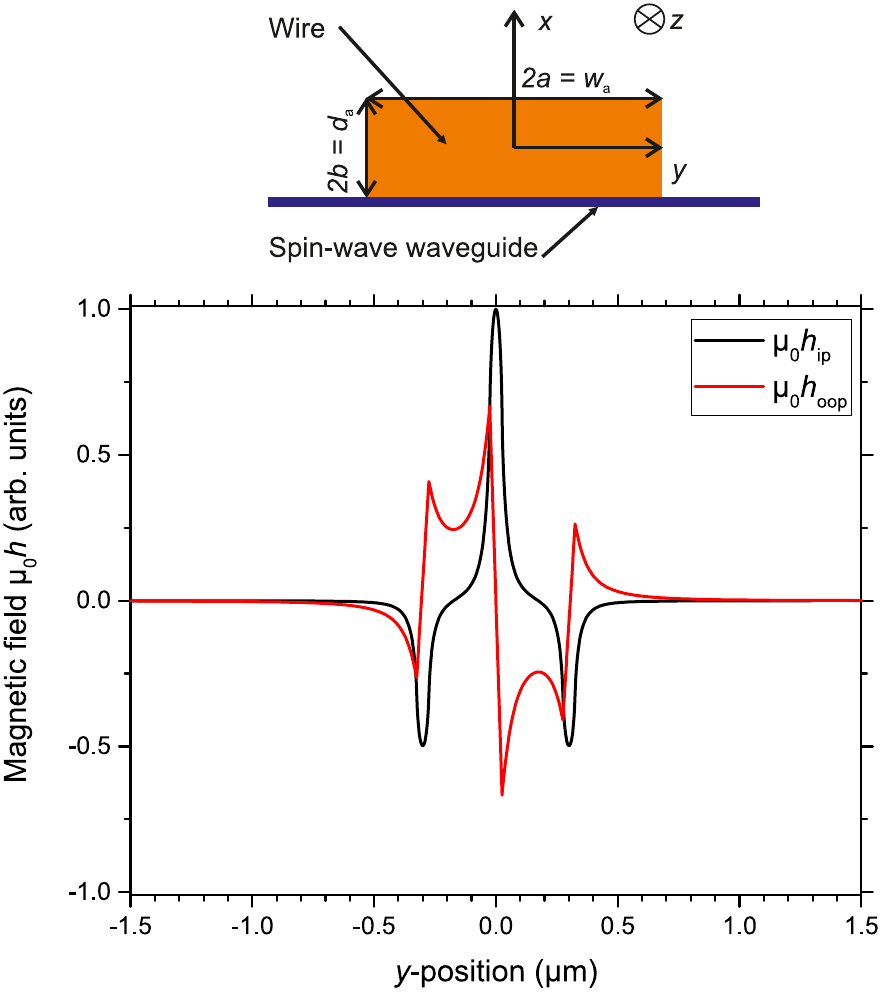}}
    \end{center}
	  \caption{\label{FigAp}(Color online) In-plane and out-of-plane component of the field of the CPW used in the calculations/simulations. The top panel shows the geometry used in the calculations of the field.}
\end{figure}

The Oersted field of the CPW is calculated by the addition of the Oersted fields created by a current flow through the three individual wires constituting the CPW. Hereby, it is assumed that the ground lines are out-of-phase with the central wire and carry each one half of the current carried in the central wire. Assuming the $z$-direction along the symmetry axis of the antenna, the in-plane field component created by an individual wire is given by\cite{Chumakov}: 
\begin{widetext}
\begin{eqnarray}
	\label{Eq:By}
	\centering
  \mu_0h_\mathrm{ip}(y,x)&=&-\frac{I \mu_\mathrm{0}}{8\pi ab}\biggl[(a-y)\biggl[\frac{1}{2}\mathrm{ln}\left(\frac{(b-x)^2+(a-y)^2}{(-b-x)^2+(a-y)^2}\right) +\frac{b-x}{a-y}\mathrm{atan}\left(\frac{a-y}{b-x}\right)\nonumber \\
& &-\frac{-b-x}{a-y}\mathrm{atan}\left(\frac{a-y}{-b-x}\right)\biggl]-(-a-y)\biggl[\frac{1}{2}\mathrm{ln}\left(\frac{(b-x)^2+(-a-y)^2}{(-a-y)^2+(-b-x)^2}\right)\nonumber \\
& &+\frac{b-x}{-a-y}\mathrm{atan}\left(\frac{-a-y}{b-x}\right)-\frac{-b-x}{-a-y}\mathrm{atan}\left(\frac{-a-y}{-b-x}\right)\biggl]\biggl]\nonumber \\
\end{eqnarray}
and the out-of-plane field component by:
\begin{eqnarray}
	\label{Eq:Bx}
	\centering
  \mu_0h_\mathrm{oop}(y,x)&=&-\frac{I \mu_0}{8\pi ab}\biggl[(b-x)\biggl[\frac{1}{2}ln\left(\frac{(b-x)^2+(a-y)^2}{(-a-y)^2+(b-x)^2}\right) +\frac{a-y}{b-x}\mathrm{atan}\left(\frac{b-x}{a-y}\right)\nonumber \\
& &-\frac{-a-y}{b-x}\mathrm{atan}\left(\frac{b-x}{-a-y}\right)\biggl]-(-b-x)\biggl[\frac{1}{2}\mathrm{ln}\left(\frac{(a-y)^2+(-b-x)^2}{(-a-y)^2+(-b-x)^2}\right)\nonumber \\
& &+\frac{a-y}{-b-x}\mathrm{atan}\left(\frac{-b-x}{a-y}\right)-\frac{-a-y}{-b-x}\mathrm{atan}\left(\frac{-b-x}{-a-y}\right)\biggl]\biggl].\nonumber \\
\end{eqnarray}  
\end{widetext}
with $a$ being the half-height and $b$ being the half-width of the wire. The superposition of the three antenna fields to the field of the CPW is depicted in Fig. \ref{FigAp} as a function of the $y$ axis, i.e., along the waveguide. The fields have been evaluated at $x = -25\,\mathrm{nm}$, i.e., just on top of the Py/Pt.

\end{document}